\documentclass[aps,pra,twocolumn,reprint,superscriptaddress,nofootinbib]{revtex4-2}

\usepackage[svgnames,table,xcdraw]{xcolor}
\usepackage{amsmath}
\usepackage{microtype}
\usepackage{amssymb}
\usepackage{braket}
\usepackage{graphicx}
\usepackage{booktabs}
\usepackage{bm}
\usepackage{dsfont}
\usepackage{hyperref}
\usepackage[caption=false]{subfig}
\usepackage{amsthm}

\usepackage{xfrac}
\usepackage{enumitem}
\usepackage{orcidlink}
\usepackage{appendix}
\usepackage[ruled,vlined]{algorithm2e}

\hypersetup{colorlinks,linkcolor={DarkBlue},citecolor={DarkBlue},urlcolor={DarkBlue}}

\let\originalleft\left
                     \let\originalright\right
\renewcommand{\left}{\mathopen{}\mathclose\bgroup\originalleft}
\renewcommand{\right}{\aftergroup\egroup\originalright}

\usepackage{mathtools}

\DontPrintSemicolon
\SetKwInput{KwInput}{Input}
\SetKwInput{KwOutput}{Output}
\SetKwFor{ForEach}{foreach}{do}{}
\SetKwFor{ForLoop}{for}{do}{}

\usepackage{nag}
\usepackage{graphicx}
\usepackage{amsthm}
\usepackage{tabulary}
\usepackage{qcircuit}
\usepackage{mathtools}
\usepackage{bm}
\usepackage{braket}
\usepackage{datetime}
\usepackage{stackrel}
\usepackage{dsfont}
\usepackage{qcircuit}
\usepackage[english]{babel}
\usepackage{units}

\usepackage{tikz}
\usetikzlibrary{backgrounds,decorations.pathreplacing}

\usepackage{chngcntr}
\counterwithout{equation}{section}

\usepackage[normalem]{ulem}
\usepackage{slashed}
\usepackage{soul}

\usepackage{xcolor}
\usepackage{amssymb}
\usepackage{amsmath}
\usepackage{amsthm}
\usepackage{braket}
\usepackage{mathdots}

\usepackage{makeidx}
\makeindex

\usepackage{soul}
\usepackage{xargs}
\newcommandx{\cmnote}[2][1=]{\linespread{1.0}\todo[linecolor=red,backgroundcolor=red!25,bordercolor=red,#1]{#2}}

\let\underline\ul

\allowdisplaybreaks[4]

 \index{} \index{} \index{} \index{} \index{} \index{} \index{} \index{}%%
\makeatletter

\newcommand{\ringplus}{\mathbin{\text{\@ringplus}}}

\newcommand{\@ringplus}{%
  \ooalign{\hidewidth\raise1.3ex\hbox{\tiny$\circ$}\hidewidth\cr$\m@th+$\cr}%
}

\newcommand{\ringminus}{\mathbin{\text{\@ringminus}}}

\newcommand{\@ringminus}{%
  \ooalign{\hidewidth\raise0.9ex\hbox{\tiny$\circ$}\hidewidth\cr$\m@th-$\cr}%
}
\makeatother
 \index{} \index{} \index{} \index{} \index{} \index{} \index{} \index{}%%

\DeclareFontFamily{U}{wncy}{}
\DeclareFontShape{U}{wncy}{m}{n}{<->wncyr10}{}
\DeclareSymbolFont{mcy}{U}{wncy}{m}{n}
\DeclareMathSymbol{\Sh}{\mathord}{mcy}{"58}

 \index{} \index{} \index{} \index{} \index{} \index{} \index{} \index{}%%
 \index{} \index{} \index{} \index{} \index{} \index{} \index{} \index{}%%

 \index{} \index{} \index{} \index{} \index{} \index{} \index{} \index{}%%
 \index{} \index{} \index{} \index{} \index{} \index{} \index{} \index{}%%
\usepackage{graphicx}
\usepackage{graphicx}
\usepackage{multirow}
\usepackage{hhline}
\xyoption{color}
\xyoption{line}

\newcommandx*\bsbal[3][1=black, 3=->]{\ar @[#1]@{#3} [#2,0] \qw}

\newcommandx*\varbs[5][1=black, 3=\theta,4=0.5,5=->]{\ar @[#1]@{#5}^(#4){#3} [#2,0] \qw}

\newcommandx*\lblline[3][3=0.5]{\ar @{-}^(#3){#1} [#2,0]}
\newcommandx*\ctrlg[3][3=0.5]{ \raisebox{-3pt}{$\bullet$}  \ar @{-}^(#3){#1} [#2,0] \qw }
\newcommandx*\ctrlog[2]{\controlo \ar @{-}^{#1} [#2,0] \qw}
\newcommandx*\ctrlodash[1]{\controlo \ar @{-} [#1,0] \ar @[black]@{.} [0,-1]}

\begin{document}

    \title{%
        \texorpdfstring
        {Decoder Dependence in Surface-Code Threshold Estimation with Native Gottesman-Kitaev-Preskill Digitization and Parallelized Sampling}
        {Decoder Dependence in Surface-Code Threshold Estimation with Native Gottesman-Kitaev-Preskill Digitization and Parallelized Sampling}
    }
    \def \affGatech {College of Computing, Georgia Institute of Technology, Atlanta, GA 30332 USA}
    \def \schrosim {Independent Quantum Architecture Researcher \& Software Developer, SchroSIM Quantum Software Project}
    \def \vgg {Volkswagen AG, Berliner Ring 2, Wolfsburg 38440, Germany}
    \def \rwth {Department of Physics, RWTH Aachen, Germany}
    \def \fujf {Department of Computational and Applied Mechanics, Federal University of Juiz de Fora, Juiz de Fora, 36036-900, Brazil}
    \def \lubeck {Institute of Information Systems, University of Lübeck, 23562 Lübeck, Germany}

    \author{Dennis Delali Kwesi Wayo \orcidlink{0000-0001-9980-6247}}
    \affiliation{\affGatech}
    \affiliation{\schrosim}
%    \email{dwayo3@gatech.edu}

    \author{Chinonso Onah}
    \affiliation{\vgg}
    \affiliation{\rwth}
%\email{chinonso.calistus.onah@volkswagen.de}
%    \email{calistusnonso@gmail.com}

    \author{Leonardo Goliatt, \,\orcidlink{0000-0002-2844-9470}}
    \affiliation{\fujf}
%    \email{goliatt@gmail.com}

    \author{Sven Groppe, \,\orcidlink{0000-0001-5196-1117}}
    \affiliation{\lubeck}
%    \email{sven.groppe@gmx.de}

    \date{\today}

    \begin{abstract}
        We quantify decoder dependence in surface-code threshold studies under two matched regimes: Pauli noise and native GKP-style Gaussian displacement digitization. Using LiDMaS+ v1.1.0, we benchmark MWPM, Union-Find (UF), Belief Propagation (BP), and neural-guided MWPM with fixed seeds, identical sweep grids, and unified reporting across runs 06--14. At $d=5$ and $\sigma=0.20$, MWPM and UF define the Pareto frontier, with (runtime, LER) = (1.341 s, 0.2273) and (1.332 s, 0.2303); neural-guided MWPM is slower and less accurate (1.396 s, 0.3730), and BP is dominated (7.640 s, 0.6107). Crossing-bootstrap diagnostics are stable only for MWPM, with median $\sigma^\star_{3,5}=0.10$ (1911/2000 valid) and $\sigma^\star_{5,7}=0.1375$ (1941/2000 valid), while other decoders show no valid crossing samples. Dense-window scanning over $\sigma \in [0.08,0.24]$ returns NaN crossings for all decoders, confirming estimator- and window-sensitive threshold localization. Rank-stability and effect-size bootstrap analyses reinforce ordering robustness: BP remains rank 4, neural-guided MWPM rank 3, and MWPM-UF differences are small ($\Delta_{\mathrm{MWPM-UF}}=-0.00383$, 95\% interval $[-0.0104,0.00329]$) across $\sigma \in [0.05,0.35]$. Threaded execution preserves statistical fidelity while improving throughput: $1.34\times$ speedup in Pauli mode and $1.94\times$ in native GKP mode, with mean $|\Delta\mathrm{LER}|$ $6.07\times10^{-3}$ and $5.20\times10^{-3}$, respectively. We therefore recommend estimator-conditional threshold reporting coupled to runtime-fidelity checks for reproducible hardware-facing practical future decoder benchmarking workflows.
    \end{abstract}

    \maketitle

    \section{Introduction}
    % \subsection{Background}
    Fault-tolerant quantum computing depends on reliable quantum error correction (QEC), and practical QEC depends on two coupled layers: a physical noise process and a classical decoder that interprets syndrome information. In idealized discussions, one often treats ``the threshold'' as a code property. In practice, threshold estimation is a joint property of code geometry, noise assumptions, syndrome extraction details, decoder objective, and statistical estimator. For this reason, two studies using the same code family can report materially different threshold values when they change decoder policy, fitting procedure, or sampling protocol. This dependence has been visible from foundational topological-code analyses through modern surface-code decoder studies \cite{surfacecode1998,fowler2009highthreshold,wang2011surfaceover1,fowler2012proofmatching,duclos2010fast,pymatching2021}.

    This work focuses on the surface code, which remains a dominant reference architecture because of its local stabilizer structure, compatibility with planar layouts, and well-developed decoder ecosystem \cite{fowler2012proofmatching}. The decoder landscape now spans exact and near-exact matching methods, near-linear-time alternatives, and tailored-noise variants whose ranking can change with channel assumptions \cite{duclos2010fast,pymatching2021,unionfind2017,tuckett2018ultrahigh,tuckett2020faulttolerant,bonilla2021xzzx,higgott2023improved,xu2023tailored}. In parallel, experimental progress has moved from repeated distance-three operation toward stronger logical suppression and below-threshold demonstrations, increasing the need for reproducible software-side comparison protocols \cite{zhao2022surfacecode,krinner2022repeated,takeda2022silicon,acharya2023suppressing,ni2023breakeven,google2025belowthreshold}.

    At the same time, emerging hardware programs increasingly motivate hybrid continuous-variable/discrete modeling choices, including GKP-inspired digitization paths and device-specific loss heterogeneity. These choices can alter both syndrome statistics and decoder behavior in ways that are not well captured by purely discrete baselines. The broader bosonic-QEC literature motivates this direction, from the original GKP encoding proposal to recent fault-tolerant and oscillator-control results in photonic and cat-qubit platforms \cite{gkp2001,tzitrin2021static,xu2023squeezedcat,reglade2024catcontrol,ding2025kerrcat}.

    LiDMaS+ was designed precisely to evaluate these interactions under one reproducible interface \cite{wayo2026lidmas}. The stack exposes a unified threshold harness, multiple decoders, native GKP-inspired digitization, optional per-qubit loss maps, and consistent outputs. This makes it possible to ask a stricter question than ``what is the threshold of this code'': under matched experimental controls, how much of a threshold claim is explained by decoder and estimator choices, and how much is explained by the underlying noise process.

    % \subsection{Problem Statement}
    Most decoder-comparison papers face a comparability problem: implementations are benchmarked under partially mismatched sweep grids, trial counts, fit procedures, or reporting conventions. Even when each individual experiment is technically valid, cross-decoder conclusions can be difficult to interpret because differences in setup are entangled with differences in algorithmic performance. The same issue appears when transitioning from Pauli-style noise to continuous-variable-informed models: if the sampling or reporting pipeline changes at the same time as the noise model, attribution becomes ambiguous. This comparability gap is visible when juxtaposing tailored-noise studies, architecture-constrained analyses, and hardware demonstrations that necessarily use different assumptions and reporting conventions \cite{bonilla2021xzzx,higgott2023improved,bravyi2024ldpcmemory,ha2025architectures,acharya2023suppressing,google2025belowthreshold}.

    A second problem is estimator overconfidence. Scalar threshold numbers are frequently presented without explicit crossing-validity diagnostics, bootstrap spread, or sweep-window dependence. In regimes where curves are shallow, non-monotonic, or distance-reversed over the sampled window, a single crossing estimate may be unstable or undefined. Without explicit diagnostics, these cases can be misread as robust threshold evidence. Recent studies that emphasize circuit-level effects and tailored boundary/noise structure reinforce the need to present threshold statements with explicit uncertainty and model conditions \cite{higgott2023improved,xu2023tailored}.

    A third problem is operational reproducibility. Large trial budgets are necessary for stable comparisons, but runtime constraints often push studies toward reduced sampling. Parallelized simulation can help, but only if serial-vs-parallel statistical agreement is checked directly. Otherwise, throughput gains may come at the cost of subtle estimator drift. This is consistent with the broader parallel-decoding literature, which treats scalability and correctness as co-requirements \cite{skoric2023parallel,tan2023scalable}.

    % \subsection{Motivation}
    The motivation for this paper is therefore methodological and practical. Methodologically, we want a decoder-comparison workflow where every major degree of freedom is explicit: noise mode, sweep design, trial budget, estimator, and post-processing logic. Practically, we want a workflow that can scale to paper-grade trial counts and still preserve reproducibility across reruns, systems, and collaborators. This follows the broader direction in decoder engineering where algorithmic speedups are meaningful only under controlled, comparable evaluation pipelines \cite{duclos2010fast,pymatching2021,unionfind2017,higgott2023improved}.

    The immediate application context is native GKP-informed surface-code simulation with additional gate/measurement/idle/loss controls, where decoder behavior can differ substantially from discrete Pauli expectations. In such settings, it is not enough to rank decoders by one curve snapshot. We need to evaluate runtime--accuracy trade-offs, stability across distances, sensitivity to channel components, effect sizes between decoders, and estimator robustness near candidate crossings. We also need to quantify whether parallelized sampling changes scientific conclusions or only reduces wall-clock time. This is aligned with ongoing work that connects bosonic and hybrid-encoding noise models to practical fault-tolerance analysis \cite{gkp2001,tzitrin2021static,xu2023squeezedcat,sahay2025transversal}.

    This is especially relevant for hardware-facing decision making. A useful decoder benchmark for device teams must answer three questions simultaneously: (i) which decoder is best in the operating regime, (ii) how reliable is that conclusion under finite sampling and estimator uncertainty, and (iii) what is the computational cost of producing the evidence. Our motivation is to provide this combined view in a form that can be regenerated from scripts and raw outputs.

    %  \subsection{Objectives}
    The primary objective is to produce a reproducible and internally consistent decoder-comparison study for surface-code threshold estimation under both Pauli and native GKP-informed noise. ``Internally consistent'' here means matched sweeps, matched trial budgets within each analysis block, deterministic seed control, and standardized output schemas.

    The secondary objective is to separate \emph{ordering claims} from \emph{threshold-scalar claims}. Ordering claims address which decoder performs better under a specified regime and metric; threshold-scalar claims assign a crossing-like critical value. We aim to show which of these claim types is robust under our diagnostics and which remains estimator-conditional.

    The tertiary objective is operational: demonstrate that higher-throughput execution (CPU threading and available acceleration paths) can support full-quality analyses while preserving close agreement with serial baselines. This objective is important because many of the diagnostics required for robust interpretation (bootstrap summaries, dense windows, ablations) are computationally expensive.

    A final objective is communication quality: translate technical decoder-benchmark results into decision-ready claims with explicit assumptions, uncertainty language, and reproducible provenance. This objective matters because practical adoption depends not only on numerical results, but also on whether independent readers can follow how each claim is produced, bounded, and regenerated from scripts.

%    \subsection{Research Questions}
    The study is guided by four questions:
    \begin{enumerate}[leftmargin=*]
    \item Decoder dependence: How do MWPM, Union-Find, BP, and neural-guided MWPM compare under matched sweeps in both Pauli and native GKP digitization modes?
    \item Stability across distance and loss: Do decoder rankings and logical-error trends remain consistent across code distances and in the presence of per-qubit loss maps?
    \item Estimator sensitivity: How sensitive are threshold summaries to the chosen estimator (crossing-based summaries versus finite-size fits) under identical simulation data?
    \item Parallelization fidelity and throughput: Does parallelized sampling (CPU threading and optional GPU offload in Pauli mode) preserve statistical equivalence of LER/CI estimates while reducing wall-clock runtime for fixed seeds and trial budgets?
    \end{enumerate}

    \subsection{Contributions}
    This paper makes the following contributions within one unified experimental pipeline:
    \begin{itemize}
        \item We provide a script-driven benchmark workflow in LiDMaS+ with matched controls across decoders and noise modes, deterministic seeding, and explicit analysis artifacts.
        \item Decoder-comparison evidence beyond single curves. We report runtime--accuracy Pareto behavior, rank stability, pairwise effect-size intervals, distance-gain diagnostics, and one-factor noise ablations in addition to baseline LER curves.
        \item Estimator-conditional threshold analysis. We report crossing/bootstrap outputs with explicit validity outcomes (including no-crossing/NaN cases) to prevent overinterpretation of scalar threshold values.
        \item Parallelization validation. We quantify speedup and serial-vs-threaded agreement under matched settings, showing how throughput can be improved without materially changing headline statistical conclusions.
        \item Implementation-aligned mathematical appendix. We include compact derivations and estimator definitions aligned with the code path used in this paper, improving traceability between study claims and executable workflow.
    \end{itemize}

    \section{Methodology}
    \label{sec:methodology}
    The implementation-aligned mathematical definitions used in this study are summarized in Appendix~\ref{app:paper02_math}. This section describes the operational protocol used to generate all reported results, with emphasis on comparability across decoders and reproducibility across reruns.

    \subsection{Study Configuration}
All experiments were executed through scripted workflows in \texttt{examples/paper\_runs/paper\_02}, with each script producing raw tabular outputs, merged datasets, and analysis-ready summaries. The experimental pipeline is organized into baseline comparisons (Pauli and native GKP fixed-distance), multi-distance sweeps, threshold-oriented analyses, parallelization tests, and advanced diagnostics (Pareto trade-off, crossing bootstrap, distance-gain heatmaps, noise ablation, rank stability, effect-size bootstrap, threading-fidelity checks, and dense critical-window sweeps). This structure enforces a one-script/one-analysis mapping and avoids hidden manual processing steps.
    Unless explicitly noted otherwise, the simulations in this paper were generated with LiDMaS+ release \texttt{v1.1.0} (commit \texttt{a1ec85e}).

    To keep comparisons fair, decoder runs share a common invocation interface and are executed with matched distance sets, sweep grids, and trial budgets for each analysis block. Where one analysis requires specialized settings (e.g., dense-window scans or one-factor ablations), those settings are applied uniformly to all decoders in that analysis. The same reporting schema is used throughout: per-point logical error rate (LER), confidence interval bounds, defect/correction diagnostics, and decoder-failure diagnostics. By construction, all derived tables and plots can be regenerated from raw tabular outputs without modifying simulation code paths.

    \subsection{Noise Models}
    We evaluate two noise regimes to separate decoder behavior under discrete and continuous-variable-inspired error channels.

    Pauli mode: The Pauli baseline samples independent Bernoulli flips on data qubits with physical error rate $p$. This mode is used as a calibration baseline because its behavior is well understood and provides a direct reference for decoder ranking and runtime comparisons under purely discrete noise.

    Native GKP mode: The native GKP path samples Gaussian displacements with standard deviation $\sigma$, applies digitization to binary flip events, and includes additional channel parameters for gate, measurement, idle, and loss processes. Per-qubit loss maps are supported through a configurable probability map. In this study, the locked default profile for native GKP analyses was
    \[
        \begin{aligned}
            \sigma &\in [0.05,0.35], \quad \Delta \sigma = 0.05, \\
            p_{\mathrm{gate}} &= 0.005, \quad p_{\mathrm{meas}} = 0.01, \\
            p_{\mathrm{idle}} &= 0.005, \quad p_{\mathrm{loss}} = 0.005.
        \end{aligned}
    \]
    except where analysis-specific overrides were required (e.g., the dense critical window in run 14).

    \subsection{Decoder Stack and Invocation Policy}
    We benchmark four decoders through the same surface-threshold interface: MWPM, Union-Find (UF), Belief Propagation (BP), and neural-guided MWPM. A key methodological choice is that the decoder list is resolved once per script and then applied uniformly across all sweep points in that script. This prevents accidental per-point decoder mismatches and ensures each reported comparison reflects a true like-for-like comparison.

    For neural-guided MWPM, the trained model path is explicitly resolved before execution; if a model is unavailable, the decoder is skipped with a warning rather than silently replaced. In the full-quality runs reported here, the reference model was present and neural-guided results are included in all relevant analyses.

    \subsection{Sweep Design and Parameter Grids}
    We use two families of sweep variables: $p$ in Pauli mode and $\sigma$ in native GKP mode. Core threshold-oriented analyses use distance set
    \[
        d\in\{3,5,7\},
    \]
    while fixed-distance comparisons use $d=5$.

    For full-quality execution, the following high-level budgets were used:
    \begin{itemize}
        \item baseline and multi-distance native GKP analyses: 3000 trials per sweep point,
        \item parallelization throughput/fidelity study: 5000 trials per sweep point,
        \item crossing bootstrap: 2000 bootstrap resamples,
        \item rank-stability bootstrap: 2000 bootstrap resamples,
        \item pairwise effect-size bootstrap: 3000 bootstrap resamples,
        \item noise ablation: 2000 trials per level with levels $\{0,0.0025,0.005,0.01\}$ for each noise component,
        \item dense critical-window sweep: $\sigma\in[0.08,0.24]$ with $\Delta\sigma=0.01$ and 2500 trials per point.
    \end{itemize}
    This mixed budget design balances precision and runtime: heavier trial counts are concentrated where conclusions are most sensitive to finite-sample variation.

    \subsection{Trial Execution, Seeding, and Statistical Endpoints}
    Each sweep point is evaluated by repeated Monte Carlo trials. Given $N$ trials and $k$ logical failures, the point estimate is
    \begin{equation}
        \widehat{\mathrm{LER}}=\frac{k}{N}.
    \end{equation}
    Confidence intervals are reported using Wilson 95\% bounds, as defined in Appendix~\ref{app:paper02_math}. In addition to LER and CI, we record:
    \begin{itemize}
        \item mean defect count,
        \item mean correction-weight proxy,
        \item decoder-failure rate.
    \end{itemize}
    These diagnostics are important for interpreting whether two decoders with similar mean LER differ in correction complexity or failure behavior.

    Reproducibility is enforced via deterministic seed mixing from base seed, distance, sweep coordinate, and trial index (with thread-aware separation). Practically, this means rerunning the same script with the same parameters reproduces the same tabular outputs, enabling exact regeneration of downstream results.

    \subsection{Experimental Protocol}
    For each decoder, distance, and sweep coordinate, we sample noise, extract syndrome(s), decode corrections, evaluate trial logical-failure indicators, and aggregate pointwise statistics. Algorithm~\ref{alg:threshold_workflow} summarizes the implementation-aligned workflow used across all reported runs.

    \begin{algorithm}[t]
    \caption{LiDMaS+ Threshold Evaluation Workflow}
    \label{alg:threshold_workflow}
    \KwInput{Mode $m\in\{\mathrm{pauli},\mathrm{gkp}\}$, decoder $\mathcal{D}$, distance set $\mathcal{D}_{\mathrm{set}}$, sweep grid $\mathcal{X}$, trials per point $N$, base seed $s_0$}
    \KwOutput{Table of $(d,x,\widehat{\mathrm{LER}},\mathrm{CI}_{95\%},\text{diagnostics})$}

    \ForEach{$d \in \mathcal{D}_{\mathrm{set}}$}{
        \ForEach{$x \in \mathcal{X}$}{
            $k \leftarrow 0$\;
            initialize defect, weight, and decoder-failure counters\;

            \ForLoop{$t \leftarrow 1$ \KwTo $N$}{
                sample physical errors with deterministic seed mix $(s_0,d,x,t)$\;
                extract syndrome(s) from sampled errors\;
                decode with $\mathcal{D}$ to correction(s)\;
                evaluate trial logical-failure indicator $\mathcal{L}_t$\;
                $k \leftarrow k + \mathcal{L}_t$\;
                update diagnostics\;
            }

            compute $\widehat{\mathrm{LER}}=k/N$ and Wilson $\mathrm{CI}_{95\%}$\;
            append one output row for $(m,\mathcal{D},d,x)$\;
        }
    }
    \end{algorithm}

    \subsection{Parallelization Metrics}
    To evaluate parallelization, we compare serial and parallelized runs under identical decoder, distance, sweep grid, trial budget, and seed schedule. In Pauli mode, the parallelized path includes CPU threading and optional GPU-accelerated sampling, while decoding remains on CPU. We report wall-clock runtime per sweep and define throughput as
    \begin{equation}
        \mathrm{throughput}=\frac{\text{total trials}}{\text{runtime (s)}}.
    \end{equation}
    Speedup is reported relative to a serial baseline:
    \begin{equation}
        S=\frac{T_{\mathrm{serial}}}{T_{\mathrm{parallel}}}.
    \end{equation}
    Statistical fidelity is evaluated pointwise on the sweep grid using LER and confidence-interval overlap. We report
    \begin{equation}
        \Delta_{\mathrm{LER}}(x)=\left|\widehat{\mathrm{LER}}_{\mathrm{serial}}(x)-\widehat{\mathrm{LER}}_{\mathrm{parallel}}(x)\right|,
    \end{equation}
    and summarize the maximum and mean $\Delta_{\mathrm{LER}}$ across sweep points. We additionally track decoder-failure-rate differences to ensure acceleration does not introduce mode-specific instability.

    \subsection{Quality Control and Validity Checks}
    We apply several consistency checks before accepting run outputs for study reporting:
    (i) sweep-grid alignment across compared decoders,
    (ii) successful output generation for each expected decoder/analysis pair,
    (iii) non-empty merged datasets before plotting,
    and (iv) explicit handling of missing crossing estimates.
    For neural-guided runs, model availability is validated before invocation. For parallelization analyses, serial and threaded outputs are compared only on shared sweep points to avoid inflated agreement metrics from unmatched coordinates.

    Finally, we separate two claim types in interpretation: \emph{ordering claims} (decoder A outperforms decoder B under a specified regime) and \emph{threshold-scalar claims} (a single crossing estimate). The former is generally robust across our diagnostics; the latter is reported conditionally and only with estimator context and crossing-validity evidence.

    \section{Results}
    We report full-quality outputs using all four decoders (MWPM, UF, BP, neural-guided MWPM) under the experimental settings described in method.~\ref{sec:methodology}. The analysis suite combines performance, stability, scaling, and reproducibility diagnostics across runs 06--14.

    \subsection{Decoder Trade-off: Accuracy Versus Runtime}
\begin{figure}[t]
        \centering
        \includegraphics[width=\columnwidth]{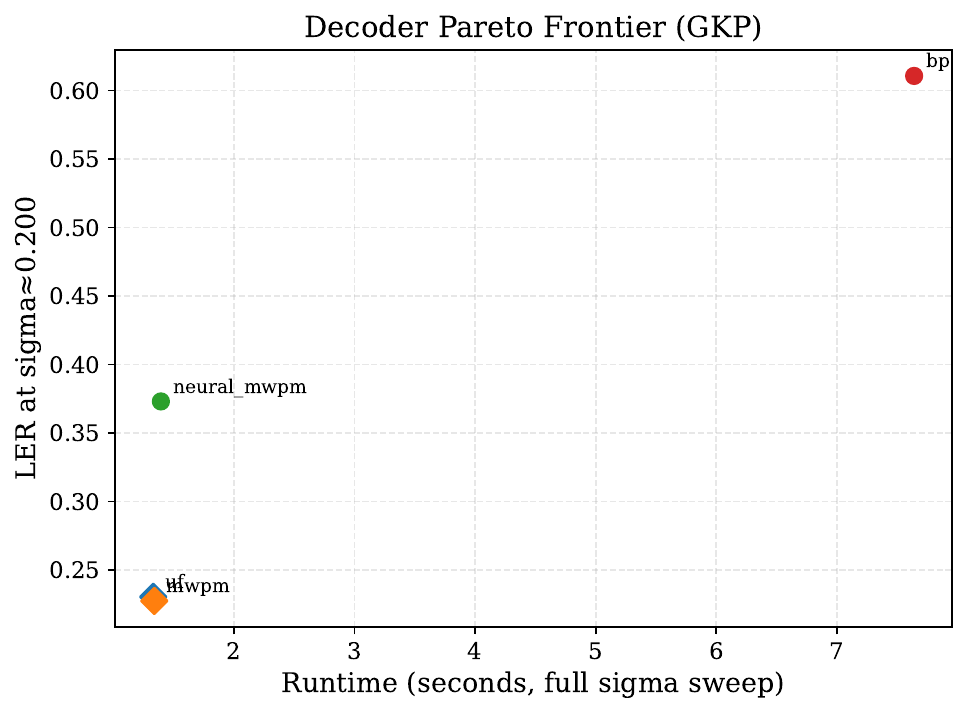}
        \caption{Decoder Pareto frontier in native GKP mode at $d=5$. Points show runtime for a full $\sigma$ sweep versus LER at $\sigma=0.20$.}
        \label{fig:pareto_decoder}
    \end{figure}

    Figure~\ref{fig:pareto_decoder} and Table~\ref{tab:pareto_decoder} summarize the runtime--accuracy frontier from the native GKP baseline sweep at fixed distance $d=5$. MWPM and UF lie on the Pareto frontier, with nearly identical runtime and logical error at $\sigma=0.20$:
    MWPM $(1.341~\mathrm{s},\,0.2273)$ and UF $(1.332~\mathrm{s},\,0.2303)$.
    Neural-guided MWPM is only slightly slower $(1.396~\mathrm{s})$ but with substantially higher error $(0.3730)$, while BP is strongly dominated by both runtime and error $(7.640~\mathrm{s},\,0.6107)$.
    This establishes a clear practical split: MWPM/UF as fast-accurate baselines, neural-guided MWPM as intermediate, and BP as high-cost/high-error in the tested regime.

    \begin{table}[t]
        \centering
        \caption{Pareto summary from native GKP baseline analysis.}
        \label{tab:pareto_decoder}
        \begin{tabular}{lccc}
            \toprule
            Decoder & Runtime (s) & LER at $\sigma=0.20$ & Pareto \\
            \midrule
            UF & 1.332 & 0.2303 & Yes \\
            MWPM & 1.341 & 0.2273 & Yes \\
            Neural-MWPM & 1.396 & 0.3730 & No \\
            BP & 7.640 & 0.6107 & No \\
            \bottomrule
        \end{tabular}
    \end{table}

    \subsection{Crossing Stability and Distance-Gain Diagnostics}
\begin{figure}[t]
        \centering
        \includegraphics[width=\columnwidth]{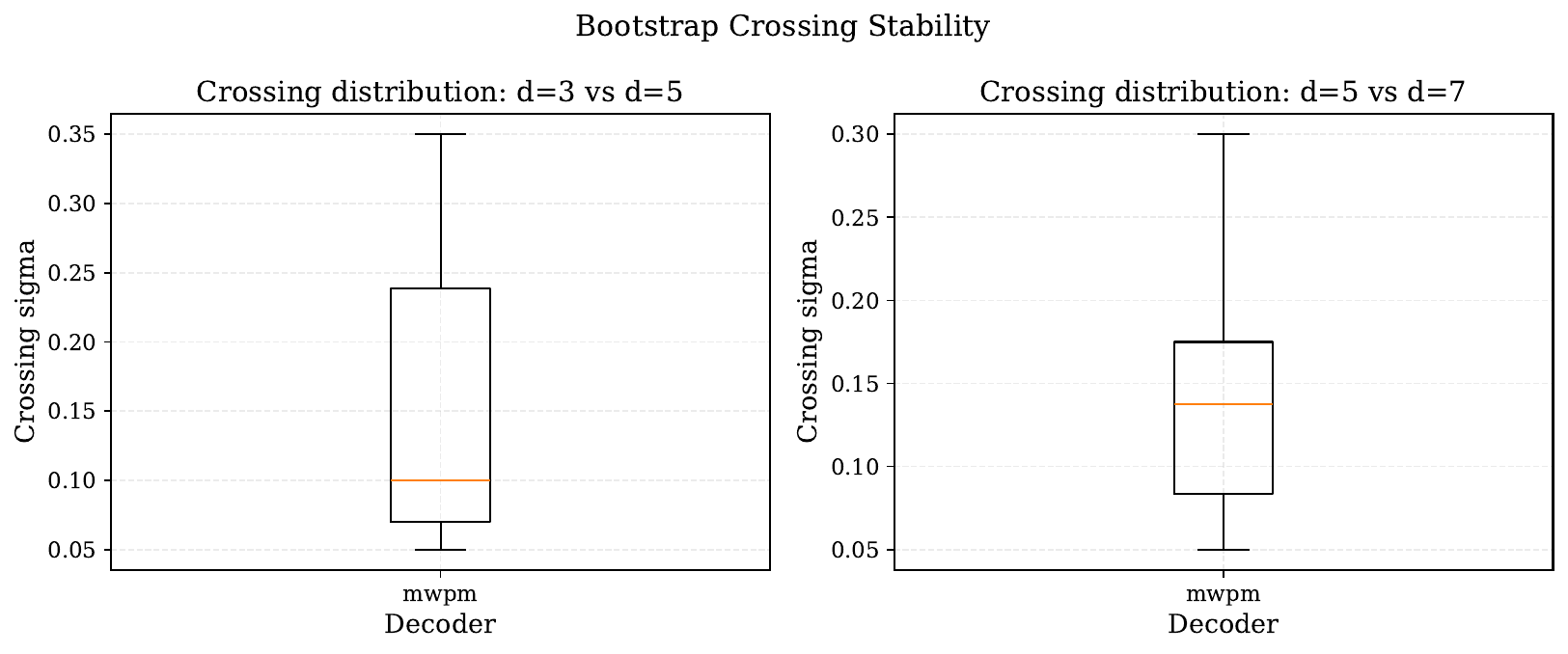}
        \caption{Bootstrap crossing distributions for $d=3/5$ and $d=5/7$ pairs. Only MWPM produced valid crossing distributions under the tested setup.}
        \label{fig:crossing_bootstrap}
    \end{figure}

    Bootstrap crossing analysis (Fig.~\ref{fig:crossing_bootstrap}) reveals that only MWPM produced non-empty crossing distributions in the sampled configuration. For MWPM, median crossings were $\sigma^\star_{3,5}=0.10$ (5--95\%: $[0.05,\,0.3167]$, 1911/2000 valid bootstrap samples) and $\sigma^\star_{5,7}=0.1375$ (5--95\%: $[0.05,\,0.20]$, 1941/2000 valid). UF, BP, and neural-guided MWPM yielded no valid crossing samples in this summary, indicating no consistent sign-change behavior under the estimator and grid used here.

\begin{figure*}[t]
        \centering
        \includegraphics[width=2\columnwidth]{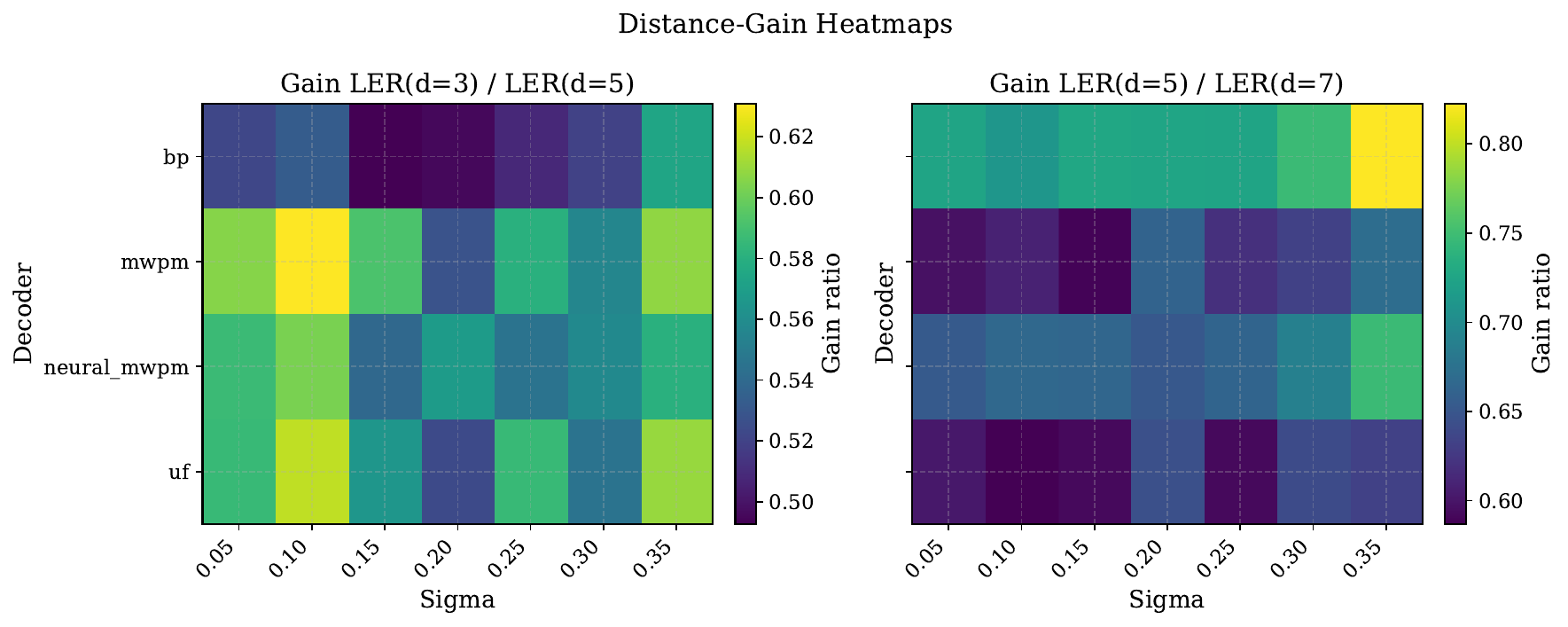}
        \caption{Distance-gain heatmaps for each decoder across $\sigma$. Values below 1 indicate that increasing code distance does not reduce LER in the sampled window.}
        \label{fig:distance_gain}
    \end{figure*}

    Distance-gain heatmaps (Fig.~\ref{fig:distance_gain}) provide complementary evidence. For all decoders and all sampled $\sigma$, the gain ratios remained below unity:
    \[
        \frac{\mathrm{LER}(d=3)}{\mathrm{LER}(d=5)} < 1, \quad
        \frac{\mathrm{LER}(d=5)}{\mathrm{LER}(d=7)} < 1.
    \]
    Mean gains were $(0.586,\,0.626)$ for MWPM, $(0.576,\,0.613)$ for UF, $(0.569,\,0.677)$ for neural-guided MWPM, and $(0.521,\,0.740)$ for BP (reported as $(d3\!\to\!d5,\ d5\!\to\!d7)$). This indicates distance reversal in the tested native GKP regime: larger distances did not yet yield lower LER on this parameter window.

    \subsection{Noise-Component Sensitivity}
\begin{figure}[t]
        \centering
        \includegraphics[width=\columnwidth]{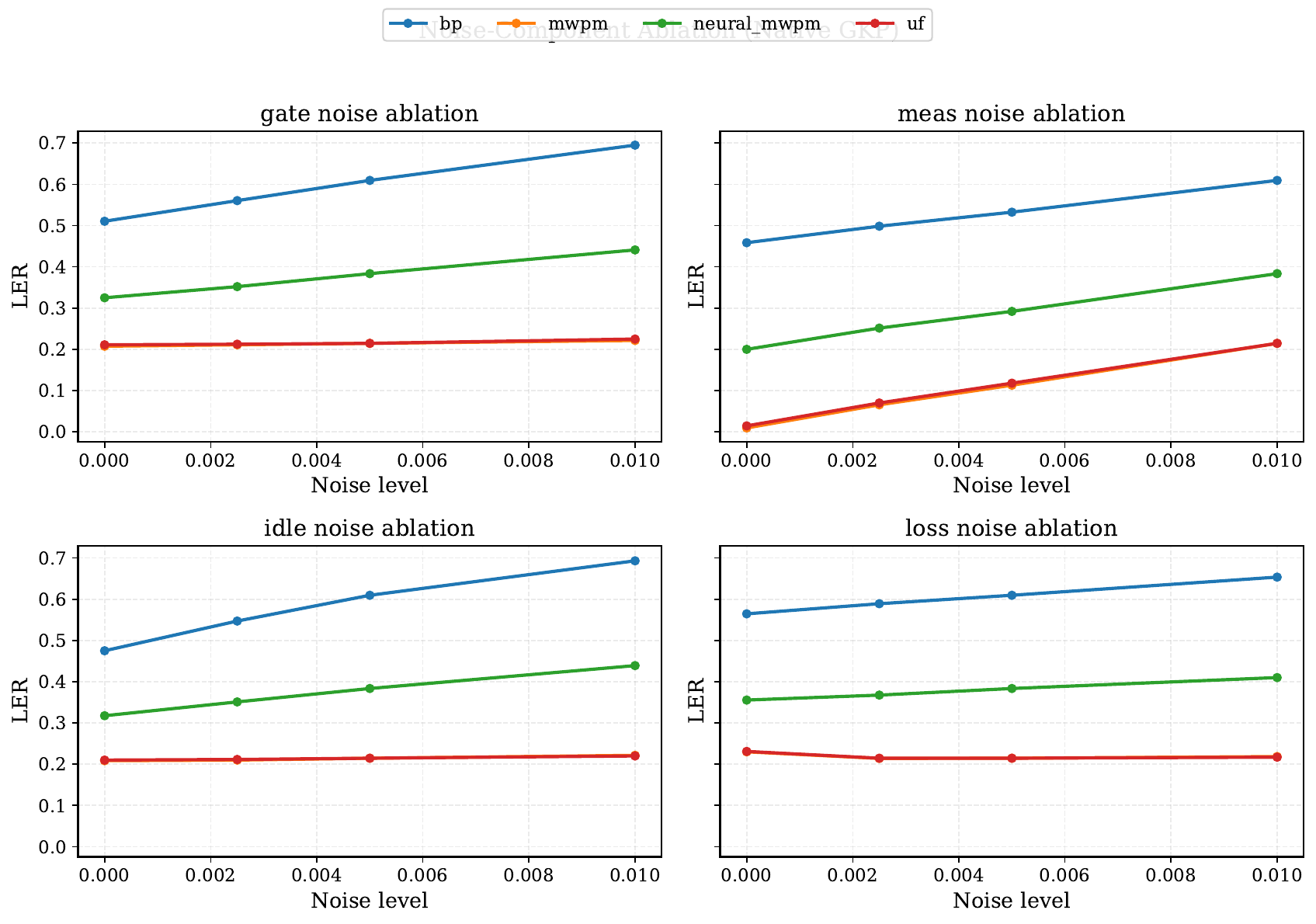}
        \caption{One-factor GKP noise ablation across gate, measurement, idle, and loss channels.}
        \label{fig:noise_ablation}
    \end{figure}

    One-factor ablation (Fig.~\ref{fig:noise_ablation}) shows measurement noise as the dominant sensitivity axis for MWPM and UF. For MWPM, the estimated slope $\partial\mathrm{LER}/\partial \epsilon$ was $20.5$ for measurement noise, versus $1.4$ (gate), $1.3$ (idle), and $-1.15$ (loss) over the tested interval $[0,0.01]$. UF follows the same pattern: $20.0$ (measurement), $1.4$ (gate), $1.0$ (idle), and $-1.4$ (loss). BP and neural-guided MWPM show broader sensitivity, with BP most sensitive to idle and gate perturbations (slopes $21.8$ and $18.45$, respectively). These results suggest that measurement-noise control is the highest-leverage axis for the MWPM/UF operating region.

    \subsection{Ranking Robustness and Pairwise Effect Sizes}
\begin{figure}[t]
        \centering
        \includegraphics[width=\columnwidth]{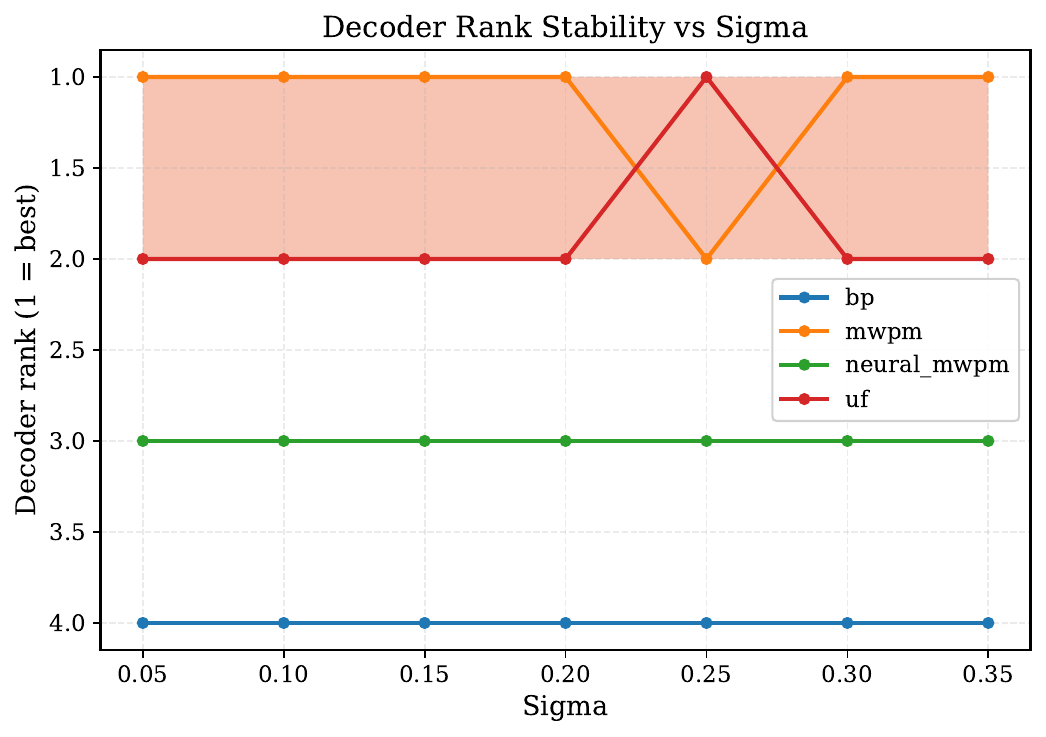}
        \caption{Decoder rank stability versus $\sigma$ with bootstrap uncertainty bands (rank 1 is best).}
        \label{fig:rank_stability}
    \end{figure}

    Rank-stability analysis (Fig.~\ref{fig:rank_stability}) shows a persistent ordering over $\sigma \in [0.05,0.35]$:
    BP stays rank 4 and neural-guided MWPM stays rank 3, while MWPM and UF alternate between ranks 1 and 2, with a local swap near $\sigma\approx0.25$.
    Bootstrap rank intervals are narrow, indicating that this ordering is not driven by high-variance sampling artifacts.

\begin{figure}[t]
        \centering
        \includegraphics[width=\columnwidth]{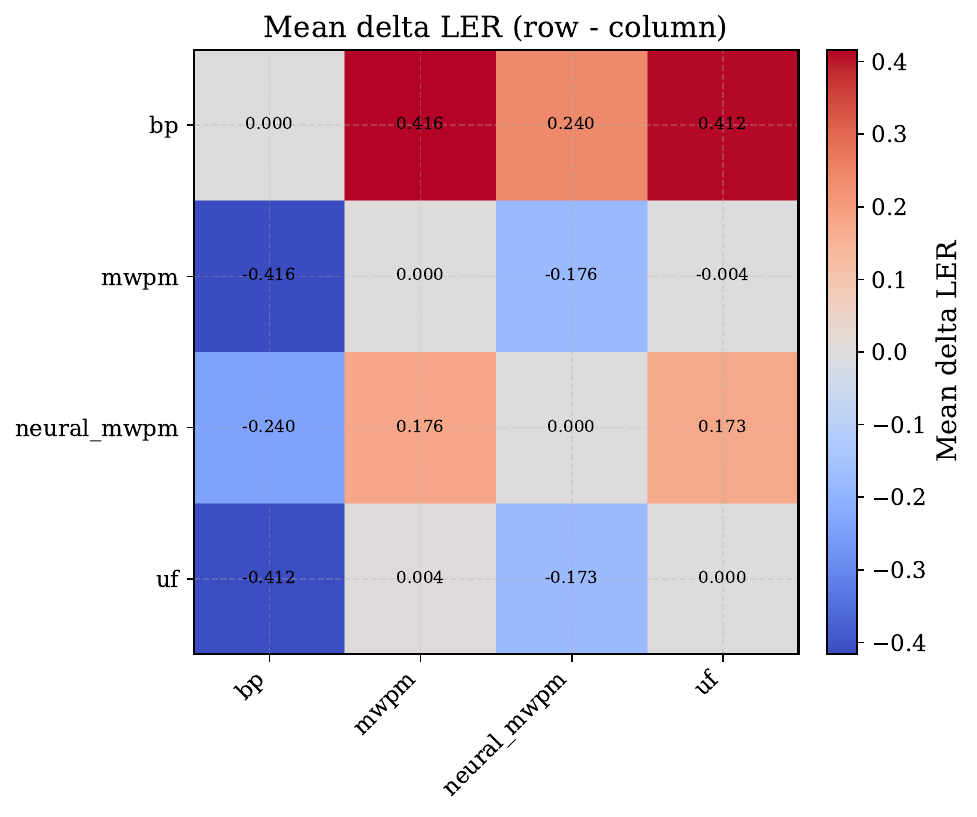}
        \caption{Pairwise mean effect-size heatmap, reported as average LER difference (row minus column).}
        \label{fig:effect_heatmap}
    \end{figure}

    Pairwise effect-size estimates (Fig.~\ref{fig:effect_heatmap}) confirm this structure quantitatively. MWPM is strongly better than neural-guided MWPM and BP:
    $\Delta_{\mathrm{MWPM-Neural}}=-0.176$ (95\% interval $[-0.183,-0.169]$),
    and by symmetry against BP (BP--MWPM mean delta $=0.416$).
    MWPM and UF are statistically close in this setup:
    $\Delta_{\mathrm{MWPM-UF}}=-0.00383$ with interval $[-0.0104,\,0.00329]$, including zero.

    \subsection{Parallelization Fidelity and Throughput}
\begin{figure*}[t]
        \centering
        \includegraphics[width=1.2\columnwidth]{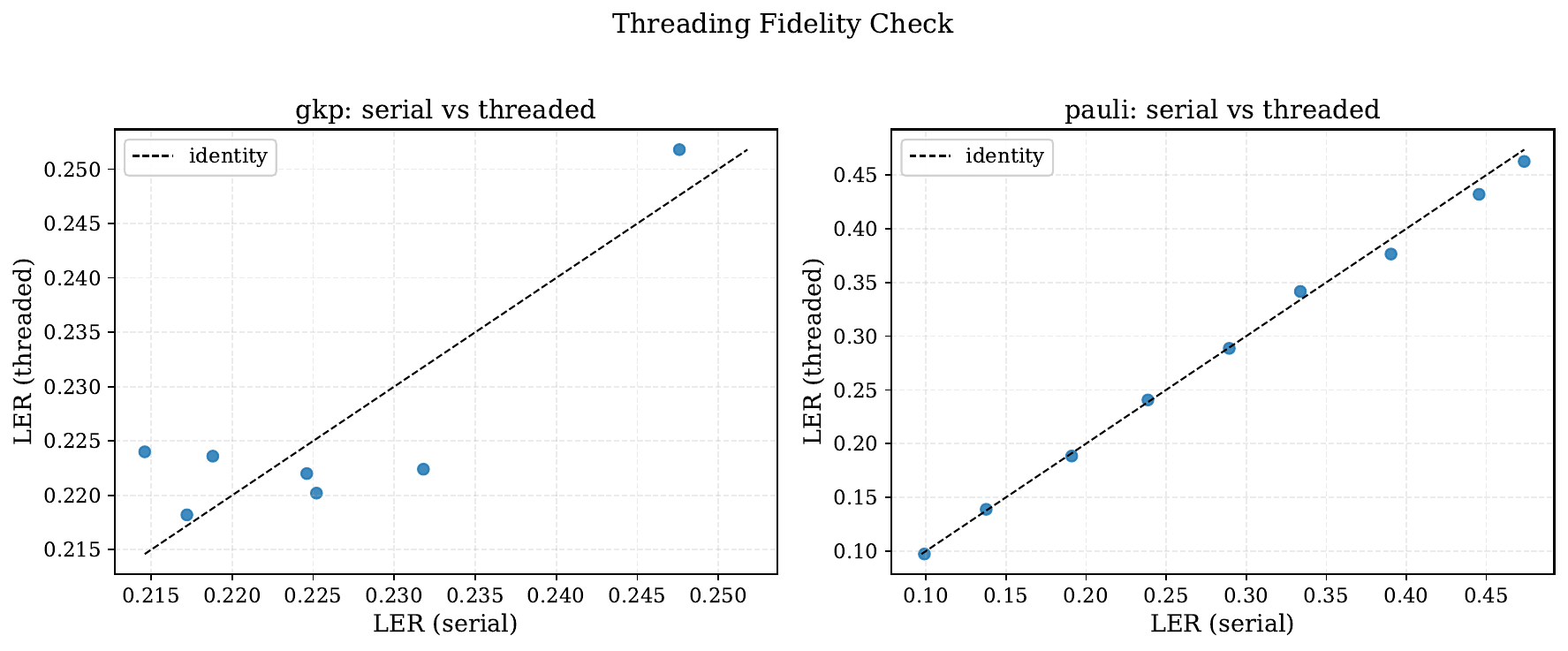}
        \caption{Serial-versus-threaded LER agreement in Pauli and native GKP modes. The dashed line denotes ideal equality.}
        \label{fig:thread_fidelity}
    \end{figure*}

    Parallelization results (Fig.~\ref{fig:thread_fidelity} and Table~\ref{tab:thread_fidelity}) show that threading accelerates sweeps while preserving close agreement in pointwise LER values. In Pauli mode, speedup is $1.34\times$ (29.76~s $\to$ 22.20~s), with mean absolute LER deviation $6.07\times 10^{-3}$ and correlation $0.9989$. In native GKP mode, speedup is $1.94\times$ (2.75~s $\to$ 1.42~s), with mean absolute LER deviation $5.2\times 10^{-3}$. The observed delta magnitudes are small relative to the absolute LER scale of the curves, supporting statistical fidelity under threaded sampling.

    \begin{table}[t]
        \centering
        \caption{Threading fidelity and throughput summary.}
        \label{tab:thread_fidelity}
        \begin{tabular}{lccc}
            \toprule
            Mode & Mean $|\Delta \mathrm{LER}|$ & Correlation & Speedup \\
            \midrule
            Pauli & 0.00607 & 0.9989 & 1.340 \\
            GKP & 0.00520 & 0.8404 & 1.940 \\
            \bottomrule
        \end{tabular}
    \end{table}

    \subsection{Dense Critical-Window Sweep}
\begin{figure}[t]
        \centering
        \includegraphics[width=\columnwidth]{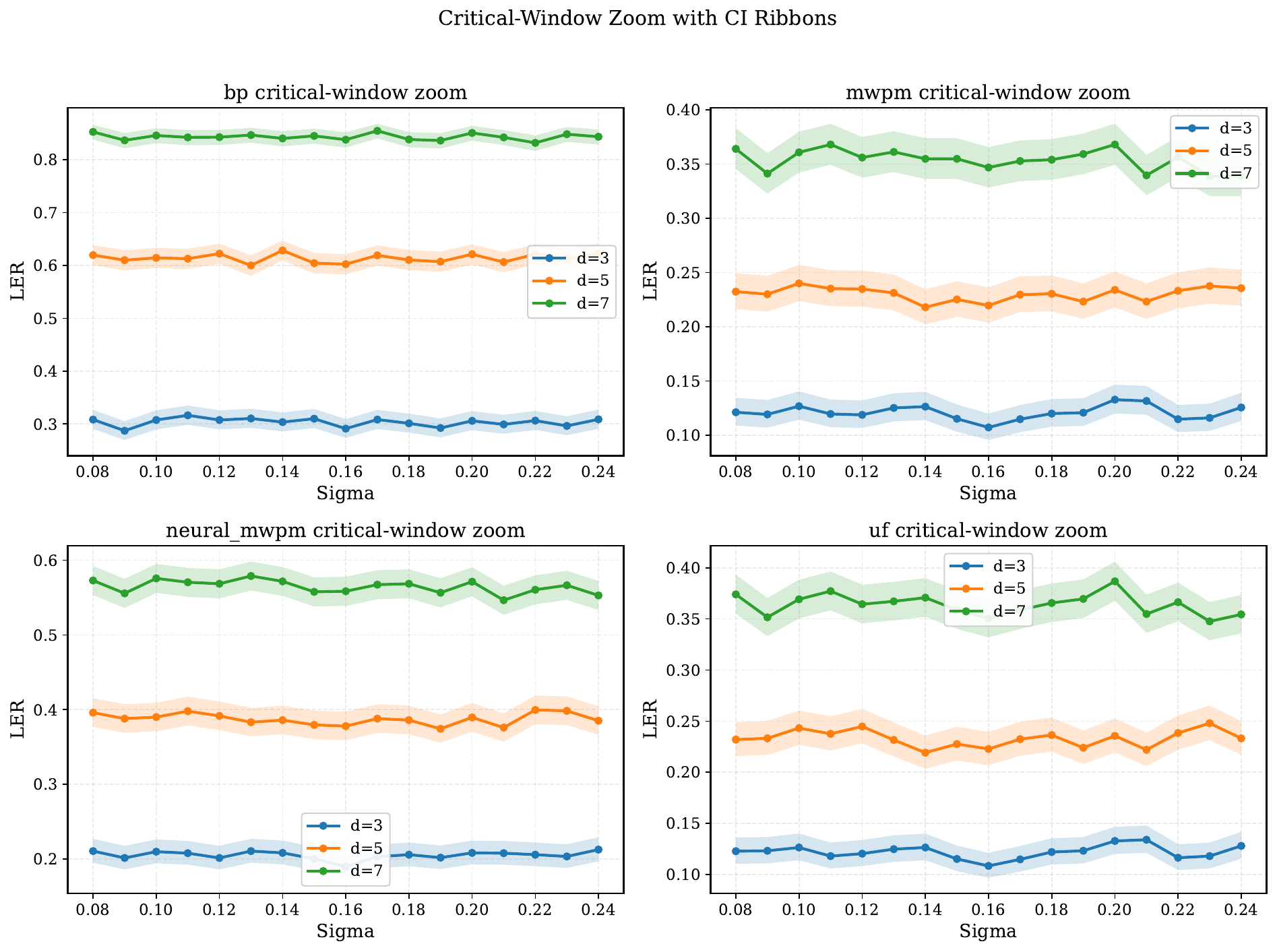}
        \caption{Dense critical-window zoom ($\sigma \in [0.08,0.24]$) with confidence ribbons for each decoder and distance.}
        \label{fig:critical_zoom}
    \end{figure}

    We then performed a dense $\sigma$ scan in the window $[0.08,0.24]$ with step $0.01$ (Fig.~\ref{fig:critical_zoom}). In this estimator pass, no decoder produced a valid crossing estimate from the combined-grid interpolation table (all entries reported as NaN in the crossing summary table). Combined with the gain-ratio and crossing-bootstrap diagnostics, this indicates that threshold localization is highly estimator- and window-sensitive in the present native GKP operating region. The data therefore support reporting practical decoder ordering and robustness diagnostics, while treating scalar threshold claims in this regime as conditional on estimator and sweep design.

    \section{Discussion}
    The main technical conclusion is that decoder dependence directly determines the strength and stability of threshold statements in native GKP-informed surface-code studies. In the tested regime, MWPM and UF remain near-tied at the top, neural-guided MWPM is consistently intermediate, and BP is consistently dominated in both error and runtime. This ordering is supported by direct LER comparisons, bootstrap rank stability, and pairwise effect-size intervals, and is broadly consistent with the practical performance reported for matching and union-find families in topological-code decoding \cite{duclos2010fast,pymatching2021,unionfind2017,higgott2023improved}.

    A second conclusion is that scalar threshold summaries are estimator-sensitive in this operating window. Crossing-bootstrap results were informative only for MWPM, dense-window interpolation summaries returned NaN crossing entries, and distance-gain ratios remained below unity across sampled $\sigma$. Together, these diagnostics indicate that threshold claims should be reported conditionally, with explicit estimator and sweep-window context. This interpretation is compatible with prior surface-code studies in which threshold location and decoder ordering vary with bias assumptions, boundary tailoring, and circuit-noise details \cite{tuckett2018ultrahigh,tuckett2020faulttolerant,bonilla2021xzzx,xu2023tailored}.

    The ablation analysis identifies measurement noise as the dominant sensitivity axis for MWPM and UF, while gate and idle channels are comparatively weaker over the tested interval. This gives a practical prioritization rule for hardware-aligned optimization: first reduce measurement-channel uncertainty, then tune lower-sensitivity channels. The recommendation is consistent with repeated-round hardware demonstrations where measurement quality and temporal stability of syndrome extraction critically affect logical suppression \cite{krinner2022repeated,acharya2023suppressing,google2025belowthreshold}.

    The parallelization study adds an operational result: threaded sampling yields substantial speedup while maintaining close statistical agreement with serial baselines. This supports production-scale sweeps with explicit serial/threaded consistency checks and aligns with prior proposals for scalable parallel decoding workflows \cite{skoric2023parallel,tan2023scalable}.

    \subsection{Implications for LiDMaS+ as a Hardware-Facing Decoder Stack}
    These findings align with the design goal of LiDMaS+ as a bridge between hardware-facing syndrome streams and decoder benchmarking. First, decoder performance should be treated as mode-dependent and configuration-dependent rather than globally ranked. Second, benchmark outputs should include uncertainty and estimator diagnostics as first-class artifacts, not appendix-only details. Third, runtime-fidelity profiling should accompany threshold plots so that decoder selection reflects both physical performance and operational cost.

    \subsection{Limitations}
    This work evaluates one code family (surface code) and one native GKP digitization pipeline. While this is sufficient to establish strong decoder dependence and estimator sensitivity in the present setting, generalization to other code families (e.g., qLDPC variants) and alternative continuous-variable mappings remains future work. In addition, the crossing behavior indicates that some parameter windows are not threshold-informative for all decoders; adaptive sweep refinement and alternative summary statistics may improve identifiability. This limitation is important given the broader diversification of code families and architectures now being explored for fault tolerance \cite{bravyi2024ldpcmemory,dua2024clifford,kobayashi2024crosscap}.

    \subsection{Recommended Reporting Practice}
    Based on the present evidence, we recommend that future threshold-comparison papers report:
    (i) decoder ranking with uncertainty bands,
    (ii) pairwise effect sizes with confidence intervals,
    (iii) crossing-validity diagnostics (including NaN or no-crossing outcomes),
    and (iv) throughput/fidelity metrics for the simulation backend.
    This reporting style prevents overinterpretation of a single threshold scalar and improves reproducibility across laboratories and software stacks.

    \section{Conclusion}
    Decoder choice, estimator design, and runtime configuration jointly determine threshold conclusions in native GKP-informed surface-code studies. Under matched conditions, MWPM and UF deliver the strongest practical trade-off, neural-guided MWPM is intermediate, and BP is dominated in the tested window. Crossing behavior is not uniformly stable across decoders, so threshold values in this regime should be presented as estimator-conditional diagnostics rather than standalone scalars.

    Measurement-channel sensitivity emerges as the key lever for the strongest decoders, and threaded sampling provides near-$2\times$ throughput gain in native GKP mode while preserving close agreement with serial baselines. These results support a reproducible, hardware-facing workflow where decoder evaluation combines error performance, uncertainty-aware threshold diagnostics, and computational efficiency.

    Next steps are to extend this protocol to additional code families and syndrome formats, incorporate adaptive sweep refinement near candidate crossings, and maintain paired serial/threaded validation as trial budgets scale. With these extensions, LiDMaS+ can serve as a robust benchmarking layer for software-comparison and hardware-in-the-loop decoder selection.

    \section*{Author Contributions}

    D.D.K.W: conceptualization, methodology, validation and visualization, software, writing – original draft, review \& editing. C.O: methodology, validation and visualization, software, writing – original draft, review \& editing. L.G: methodology, validation and visualization, software, writing – original draft, review \& editing. S.G: methodology, validation and visualization, software, writing – original draft, review \& editing.

    \section*{Acknowledgment(s)}

    The authors acknowledge contributors, users, and reviewers who provided
    feedback on decoding workflows, reproducibility scripts, and
    documentation quality. Any opinions, findings, conclusions, or recommendations expressed in this research are those of the author(s) and do not necessarily reflect the views of their respective affiliations.

    \section*{Data \& Code Availability}
    The data generated and analyzed during the present study are included within the manuscript. Supplementary codes developed from \texttt{LiDMaS+} simulator are provided as supplementary material and accessible on \href{https://github.com/DennisWayo/lidmas_cpp}{GitHub} to ensure transparency and reproducibility.

    \section*{Funding}
    This research was not funded.

    \section*{Disclosure statement}

    No potential conflict of interest was reported by the author(s).

    \appendix
    \section{Implementation-Aligned Mathematical Formulation}
    \label{app:paper02_math}

    \subsection{Binary Algebra and Syndromes}
    Let all parity operations be over $\mathbb{F}_2=\{0,1\}$. For parity-check matrices
    \begin{equation}
        H_X \in \mathbb{F}_2^{m_X\times n},
        \qquad
        H_Z \in \mathbb{F}_2^{m_Z\times n},
    \end{equation}
    and data-error vectors $e_X,e_Z\in\mathbb{F}_2^n$, the measured syndromes are
    \begin{equation}
        s_Z = H_Z e_X \bmod 2,
        \qquad
        s_X = H_X e_Z \bmod 2.
    \end{equation}
    For a correction $(c_X,c_Z)$, residual errors are
    \begin{equation}
        e_X^{\mathrm{res}}=e_X\oplus c_X,
        \qquad
        e_Z^{\mathrm{res}}=e_Z\oplus c_Z.
    \end{equation}

    \subsection{Noise-Model Parameterization}
    In Pauli mode, per-qubit flips are sampled as Bernoulli variables:
    \begin{equation}
        \Pr(e_{X,i}=1)=p,
        \qquad
        \Pr(e_{Z,i}=1)=p.
    \end{equation}

    In native GKP mode, displacement samples are
    \begin{equation}
        \Delta q,\Delta p \sim \mathcal{N}(0,\sigma^2),
    \end{equation}
    digitized on lattice scale $\lambda=\sqrt{\pi}$:
    \begin{equation}
        n_q=\mathrm{round}(\Delta q/\lambda),\quad
        n_p=\mathrm{round}(\Delta p/\lambda),
    \end{equation}
    \begin{equation}
        x_{\mathrm{flip}}=|n_q|\bmod 2,\quad
        z_{\mathrm{flip}}=|n_p|\bmod 2.
    \end{equation}
    Gate, idle, measurement, and loss channels are applied independently using
    probabilities $(p_{\mathrm{gate}},p_{\mathrm{idle}},p_{\mathrm{meas}},p_{\mathrm{loss}})$,
    with optional per-qubit loss map $p_{\mathrm{loss}}(q)$.

    \subsection{Logical-Failure Indicator and LER Estimation}
    For logical supports $(L_X,L_Z)$, define the trial failure indicator
    \begin{equation}
        \mathcal{L}_t=
        \mathbf{1}\!\left[\langle e_{X,t}^{\mathrm{res}},L_X\rangle=1
                        \ \lor\
                        \langle e_{Z,t}^{\mathrm{res}},L_Z\rangle=1\right].
    \end{equation}
    Over $N$ trials at one sweep point, with $k=\sum_{t=1}^N \mathcal{L}_t$, the logical
    error-rate estimator is
    \begin{equation}
        \widehat{\mathrm{LER}} = \frac{k}{N}.
    \end{equation}
    We report Wilson 95\% confidence intervals ($z=1.9599639845$):
    \begin{align}
        \mathrm{den} &= 1 + \frac{z^2}{N},\\
        c &= \frac{\hat p + z^2/(2N)}{\mathrm{den}},\qquad \hat p=\frac{k}{N},\\
        h &= \frac{z}{\mathrm{den}}
        \sqrt{\frac{\hat p(1-\hat p)+z^2/(4N)}{N}},
    \end{align}
    \begin{equation}
        \mathrm{CI}_{95\%} = [\max(0,c-h),\ \min(1,c+h)].
    \end{equation}

    \subsection{Crossing and Bootstrap Summaries}
    For two distances $d_1,d_2$ on sweep variable $x\in\{p,\sigma\}$, define
    \begin{equation}
        \Delta(x)=\widehat{\mathrm{LER}}_{d_1}(x)-\widehat{\mathrm{LER}}_{d_2}(x).
    \end{equation}
    If a sign change exists on $[x_k,x_{k+1}]$, the piecewise-linear crossing estimate is
    \begin{equation}
        x^\star = x_k + (x_{k+1}-x_k)\frac{\Delta(x_k)}{\Delta(x_k)-\Delta(x_{k+1})}.
    \end{equation}
    Bootstrap distributions are obtained by pointwise binomial resampling:
    \begin{equation}
        k' \sim \mathrm{Binomial}(N,\hat p),\qquad \hat p'=\frac{k'}{N},
    \end{equation}
    and recomputing $x^\star$ across resamples.

    \subsection{Pairwise Effect Size and Rank Stability}
    For decoders $a,b$, we report mean effect size over shared sweep points
    \begin{equation}
        \Delta_{a-b} = \frac{1}{M}\sum_{j=1}^{M}
        \left(\widehat{\mathrm{LER}}_a(x_j)-\widehat{\mathrm{LER}}_b(x_j)\right),
    \end{equation}
    with bootstrap confidence intervals from the same binomial-resampling procedure.
    Rank stability is computed by ranking decoder LER values at each $x_j$ and
    summarizing bootstrap quantiles of rank trajectories.

    \subsection{Parallelization Fidelity and Throughput}
    Throughput and speedup are
    \begin{equation}
        \mathrm{throughput}=\frac{N_{\mathrm{total}}}{T},
        \qquad
        S=\frac{T_{\mathrm{serial}}}{T_{\mathrm{parallel}}}.
    \end{equation}
    Serial/threaded agreement is summarized with
    \begin{equation}
        \Delta_{\mathrm{LER}}(x)=
        \left|
            \widehat{\mathrm{LER}}_{\mathrm{serial}}(x)-
            \widehat{\mathrm{LER}}_{\mathrm{threaded}}(x)
        \right|,
    \end{equation}
    and reported via mean/max values over the full sweep.

    \bibliographystyle{apsrev4-2}
    \bibliography{lidmas}

\end{document}